\documentclass[aps,11pt,showpacs,nofootinbib,preprintnumbers,floatfix]{revtex4}
\usepackage{}
\usepackage{graphicx}
\usepackage{epsfig}
\usepackage{dcolumn}
\usepackage{subfigure}
\usepackage{amsmath}

\def\be{\begin{equation}}
\def\ee{\end{equation}}
\def\bea{\begin{eqnarray}}
\def\eea{\end{eqnarray}}

\def\d#1#2{\frac{\displaystyle #1}{\displaystyle #2}}

\def\p{\partial}
\newcommand{\omits}[1]{}

\def\bsp{\be\begin{split}}

\def\bes{\be  \begin{split}}

\def\p{\partial}

\newcommand{\Rmnum}[1]{\expandafter\@slowromancap\romannumeral #1@}

\def\PRD{{Phys. Rev.}~{\bf D}}
\def\PRL{{Phys. Rev. Lett. }}
\def\PLA{{Phys. Lett.}~{\bf A}}
\def\PLB{{Phys. Lett.}~{\bf B}}
\def\GRG{{Gen. Rel. Grav. }}
\def\CQG{{Class. Quant. Grav. }}

\def\JHEP{{JHEP}}

\def\IJMPD{{Int. J. Mod. Phys.}~{\bf D}}

\begin{document}

\title{Magnetically charged regular black hole in a model of nonlinear electrodynamics}
\author{Meng-Sen Ma$^{1,2}$\footnote{Email: mengsenma@gmail.com}}

\medskip

\affiliation{\footnotesize$^1$Department of Physics, Shanxi Datong
University,  Datong 037009, China\\
\footnotesize$^1$Institute of Theoretical Physics, Shanxi Datong
University, Datong 037009, China}

\begin{abstract}
We obtain a magnetically charged regular black hole in general relativity. The source to the Einstein field equations is nonlinear electrodynamic field in a physically reasonable model of nonlinear electrodynamics (NED). ``Physically" here means the NED model is constructed on the basis of three conditions: the Maxwell asymptotic in the weak electromagnetic field limit; the presence of vacuum birefringence phenomenon; and satisfying the weak energy condition (WEC). In addition, we analyze the thermodynamic properties of the regular black hole in two ways. According to the usual black hole thermodynamics, we calculate the heat capacity at constant charge, from which we know the smaller black hole is more stable. We also employ the horizon thermodynamics to discuss the thermodynamic quantities, especially the heat capacity at constant pressure.

\end{abstract}

 \maketitle

\section{Introduction}

The first well-known model of nonlinear electrodynamics(NED) is the Born-Infeld theory(BI), which is proposed to obtain a finite electron self-energy\cite{BI:1934}.
Heisenberg and Euler found that due to the presence of virtual charged particles the one-loop quantum correction in quantum electrodynamics will give nonlinear contribution\cite{HE:1936}.
Not only that, the virtual particles will result in `` polarization of the vacuum ". In this case, the vacuum behaves like a polarizable continuum and should exhibit the phenomenon of birefringence\cite{Klein:1964,Brezin:1970,Heinzl:2006}. These effects can be observed in experiments such as PVLAS\cite{PVLAS} and BMV\cite{BMV} and
 the experimental results can put some restrictions on the parameters introduced in the NED models. Vacuum birefringence is a nonlinear effect. Therefore, we expect that in a physically reasonable model of NED, there should exist the effect of vacuum birefringence. In this sense, the usual BI theory is not a physically allowable NED model because of the absence of vacuum birefringence. However, in the generalized BI model with two parameters the vacuum birefringence may exist\cite{Kruglov:2010}. The nonlinear effects in electrodynamics are significant only for strong electromagnetic fields. In weak field case,  NED should return back to the conventional Maxwell theory, just like the weak field approximation of general relativity gives the Newtonian mechanics. Thus, we expect that this is another requirement for a physically reliable NED model.

Recently, due to the natural emergence in string theories\cite{Fradkin:1985,Callan:1987,Brecher:1998}, NED is coming into view in gravitational theories as source.
It is well-known that by minimal coupling of Maxwell electromagnetic fields to gravity, the Reissner-Nordstr\"{o}m(RN) black hole can be obtained. Similarly, more interesting black hole solutions can be derived through minimal coupling to gravity of nonlinear electromagnetic fields, such as BI and BI-like electromagnetic fields\cite{Hoffmann,Oliveira}, logarithmic electromagnetic field\cite{Soleng}, power electromagnetic field\cite{Martinez:2008} and exponential electromagnetic field\cite{Hendi:2012}.

In this letter, we shall investigate regular black hole solution derived in general relativity coupled to NED. The first example of a
regular black hole was constructed by Bardeen in 1968\cite{Bardeen}.
Nearly thirty years later, Ay\'{o}n-Beato, et al reobtained the Bardeen
black hole by describing it as the gravitational field of a kind of
nonlinear magnetic monopole\cite{RBH3}. Similarly, many other
regular black holes can also be constructed by introducing nonlinear
electromagnetic sources\cite{RBH1,RBH2,EAB:1999,Bronnikov:2001,Dymnikova:2004}. There are also other types of regular/nonsingular black holes with different origins.
For more detailed description of regular black holes, one can refer to the paper by Lemos et.al\cite{Lemos:2011} and references therein. We should stress that there is a theorem which asserts that the existence of electrically charged, static, spherically symmetric solutions with a regular center is forbidden,
while the existence of the solutions with magnetic charges is feasible, if the NED model contains the Maxwell theory as its weak approximations\cite{Bronnikov:2001}.
Thus, we only concern with the magnetically charged regular black hole with the special emphasis put on its thermodynamics and stability.

The paper is arranged as follows: in the next section we simply
introduce the NED model and its origin. In section 3  we will solve the Einstein field equations to obtain the regular black hole solution
and analyze the geometric structure. In section 4 we verify that the nonlinear electromagnetic field in this NED model satisfy the WEC.
Then we calculate the thermodynamic quantities, such as temperature, the heat capacity at constant charge and the heat capacity at constant pressure
from which we can discuss the local stability of the regular black holes in section 5.  We will make some
concluding remarks in section 6.

\par

\section{The NED model}

Firstly we simply introduce the model of NED proposed by Kruglov\cite{Kruglov:2015}, which can produce the vacuum birefringence phenomenon.
The Lagrangian is given by
\be
\mathcal{L}=\mathcal{F}+\d{a\mathcal{F}}{2\beta \mathcal{F}+1}-\d{\gamma}{2}\mathcal {G}^2,
\ee
where $a$ is a dimensionless parameter and $\beta,~\gamma$ are parameters with the dimensions of $[L^2]$.
$\mathcal{F}=F_{\mu\nu}F^{\mu\nu}, ~\mathcal {G}=F_{\mu\nu}\tilde{F}^{\mu\nu}$ are two Lorentz invariants with
$\tilde{F}_{\mu\nu}=1/2\varepsilon_{\mu\nu\alpha\beta}F^{\alpha\beta}$, and
$F_{\mu\nu}=\partial_{\mu}A_{\nu}-\partial_{\nu}A_{\mu}$ is the electromagnetic field strength.

It is shown that in a constant and uniform external magnetic field $\textbf{B}_0$, the indexes of refraction with different polarizations of electromagnetic waves are
\be
n_{\parallel}=\sqrt{1+\d{\gamma B_0^2(\beta B_0^2+1)^2}{a+(\beta B_0^2+1)^2}}, \quad n_{\perp}=1.
\ee
This means that the electromagnetic waves with different polarizations have different velocities and thus the vacuum birefringence is present.
However, one can easily see that this NED model cannot return back to the Maxwell theory in the weak approximation except for $a=0$, because
\be
\mathcal{L}\approx (1+a)\mathcal{F}+O[\mathcal{F}^2].
\ee
Therefore, below we shall consider a deformed NED model,
\be\label{Lag}
\mathcal{L}=\d{\mathcal{F}}{2\beta \mathcal{F}+1}-\d{\gamma}{2}\mathcal {G}^2.
\ee
Obviously, this model contains Maxwell theory as weak field approximation. Moreover, this model can still produce the vacuum birefringence phenomenon\footnote{Kruglov points out this fact for us.}.

\section{The magnetically charged regular black hole}

We consider the NED in the framework of general relativity, with the action
\be
S=\d{1}{16\pi}\int d^4x\sqrt{-g}\left[R-\mathcal{L(\mathcal{F},\mathcal{G})}\right],
\ee
where $\mathcal{L(\mathcal{F},\mathcal{G})}$ takes the form of Eq.(\ref{Lag}).

We only consider the static, spherically symmetric spacetime, in Schwarzschild gauge which can be written in the form
\be\label{staticmetric}
ds^2=-f(r)dt^2+f(r)^{-1}dr^2 + r^2d\Omega^2.
\ee
The dynamical equation for the electromagnetic fields is
\be
\nabla_\mu(L_{F}F^{\mu\nu}+L_{G}\tilde{F}^{\mu\nu})=0,
\ee
where $L_F=d\mathcal{L}/d\mathcal{F}, ~L_{G}=d\mathcal{L}/d\mathcal{G}$, and the corresponding Bianchi identity is
\be
\nabla_\mu\tilde{F}^{\mu\nu}=0.
\ee
According to the no-go theorem proposed by Bronnikov, to find the regular black hole solution, we will only consider the pure magnetic case\cite{Bronnikov:2001}. In this case the $\mathcal{G}$-square term in Eq.({\ref{Lag}) is zero. Thus the Lagrangian of electromagnetic fields now becomes
\be\label{Lag1}
\mathcal{L(\mathcal{F})}=\d{\mathcal{F}}{2\beta \mathcal{F}+1}.
\ee
In spherically symmetric case, $F_{\mu\nu}$ involves a radial magnetic field $F_{23}$  and satisfies
\be
F_{23}=q_m \sin\theta,
\ee
where $q_m$ is the magnetic charge. For clarity and simplicity, we write $q_m$ as $q$ and take it to be positive below. Thus, $\mathcal{F}=2F_{23}F^{23}=2q^2/r^4$.

From Eq.(\ref{Lag1}), one can easily seen that  $\mathcal{L(\mathcal{F})}\rightarrow \mathcal{F},~ L_F\rightarrow 1$ when $\mathcal{F}\rightarrow 0$. This is the Maxwell asymptotic. Particularly, $\mathcal{L(\mathcal{F})}\rightarrow 1/2\beta=const$ when $F\rightarrow \infty$, which means that near $r=0$ the energy-momentum tensor $T_{\mu\nu}$ is a constant.
According to the discussion in \cite{Bronnikov:2001}, the spacetime near $r=0$ is the de Sitter one and must be regular there.

The metric function can be written in the form
\be\label{fmr}
f(r)=1-\d{2m(r)}{r}.
\ee
Substituting it into the Einstein field equations, one can obtain
\be\label{eom}
G_{~0}^{0}=G_{~1}^{1}=-\d{2m'(r)}{r^2}=-T_{~0}^{0}=-T_{~1}^{1}=-\mathcal{L}/2.
\ee
Integrating the above equation, we have
\be\label{mr}
m(r)=M-\d{1}{4}\int_{r}^{\infty}r^2\mathcal{L}dr,
\ee
where $M$ is an integration constant and is chosen to satisfy $m(\infty)=M$. The solution (\ref{mr}) satisfies the field equations of $(_{2}^{2})$ and $(_{3}^{3})$ components automatically.

The mass function takes the form
\be\label{mr1}
m(r)=\frac{q^{3/2}}{16 \sqrt[4]{\beta }} \left[\ln \frac{2 \sqrt{\beta } q-2 \sqrt[4]{\beta } \sqrt{q} r+r^2}{2 \sqrt{\beta } q+2 \sqrt[4]{\beta } \sqrt{q} r+r^2}+
2\tan ^{-1}\left(1+\frac{r}{\sqrt[4]{\beta } \sqrt{q}}\right)-2 \tan ^{-1}\left(1-\frac{r}{\sqrt[4]{\beta } \sqrt{q}}\right)\right]
\ee
To obtain a black hole solution which is regular at $r=0$, the parameter $\beta$ must take the value
\be
\beta=\left(\frac{\pi  q^{3/2}}{8 M}\right)^4.
\ee
Thus the metric function is
\be
f(r)=1+\frac{M}{\pi  r}\left[\ln \d{32 M^2 r^2+8 \pi  M q^2 r+\pi ^2 q^4}{32 M^2 r^2-8 \pi  M q^2 r+\pi ^2 q^4}-2 \tan ^{-1}\left(\frac{8 M r}{\pi  q^2}+1\right)+2 \tan ^{-1}\left(1-\frac{8 M r}{\pi  q^2}\right)\right].
\ee
Expanding the metric function near $r=0$, one can find that there is indeed a de Sitter core,
\be
f(r)=1-\frac{1024 M^4 r^2}{3 \left(\pi ^4 q^6\right)}+O(r)^6.
\ee
And expanding the metric in the $r\to \infty$ limit, we can obtain
\be
f(r)=1-\frac{2 M}{r}+\frac{q^2}{r^2}+O(r)^6.
\ee
Obviously, it recovers the RN black hole in the $r\to \infty$ limit.

There are three variables,$(r,~q,~M)$ in the metric function. We can introduce a new coordinate
$x=r/M$ and a new charge $Q=M/q$. The parameters $M, q$ can be cancelled in the metric function, we obtain
\be
f(x)=1+ \d{1}{\pi x}\left[\ln \d{32 Q^4 x^2+8 \pi  Q^2 x+\pi ^2}{32 Q^4 x^2-8 \pi  Q^2 x+\pi ^2}+2 \tan ^{-1}\left(1-\frac{8 Q^2 x}{\pi }\right)-2 \tan ^{-1}\left(\frac{8 Q^2 x}{\pi }+1\right)  \right].
\ee
The horizons correspond to the locations where $f(x)=0$. The equation is too complicated to express the roots analytically. As is shown in Fig.\ref{fig1}, there can be one or two  horizons depending on the value of $Q$. Particularly, for certain choices of $Q$, there can be no horizon, leading to a particle-like solution. For fixed $x$, if we decrease the value of $Q$, these two horizons of the regular black hole come closer and they meet at the critical value of $Q=Q_e$. In this time, the black hole is extremal. Numerically, one has $x_e=0.9526$ and $Q_e=0.9893$. This means that only when $M\geq0.9893q$ the horizons may exist.

\begin{figure}[!htbp]
\includegraphics[width=7cm,keepaspectratio]{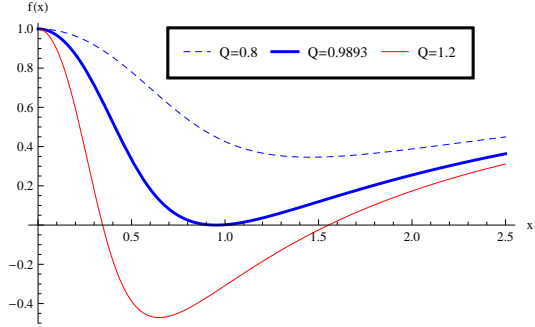}\hspace{0.5cm}
\caption{The metric function $f(x)$ as function  of  $x$. The dashed line, solid(blue, thick) line, and solid(red) line correspond to $Q=0.8, ~0.9893,~1.2$, respectively. \label{fig1}
}
\end{figure}

\section{Weak energy condition}

The weak energy condition(WEC) corresponds to the statement $T_{\mu\nu}u^{\mu}u^{\nu}\geq0$, which means that the energy density of any matter distribution as measured by any local observer in spacetime must be nonnegative and dominates over the pressure\cite{Poisson}. Except for some special cases or matter fields, the WEC is fulfilled by general matter fields.  Therefore we expect that a physically acceptable NED model cannot violate the WEC. According to the metric ansatz and Eq.(\ref{fmr}), WEC can be expressed equivalently as the following inequalities\cite{Vagenas:2014}:
\be
\d{1}{r^2}\d{d m(r)}{d r}\geq 0, \quad \d{2}{r}\d{d m(r)}{d r}\geq \d{d^2 m(r)}{d r^2}.
\ee
From Eq.(\ref{mr1}), it can be easily derived that
\be
\d{d m(r)}{d r}=\frac{q^2 r^2}{8 \beta  q^2+2 r^4},
\ee
and
\be
\d{2}{r}\d{d m(r)}{d r}-\d{d^2 m(r)}{d r^2}=\frac{2 q^2 r^5}{\left(4 \beta  q^2+r^4\right)^2}.
\ee
Clearly, the NED model we considered satisfies the WEC.

\section{Thermodynamics}

We now would like to study the thermodynamic properties of the regular black hole we obtained.

The Hawking temperature of the black hole can be calculated using the formula
\be\label{tem}
T_h=\d{\kappa}{2\pi}=-\left.\d{1}{4\pi}\d{\p_{r}g_{tt}}{\sqrt{-g_{tt}g_{rr}}}\right|_{r=r_{+}}=
\left.\d{1}{4\pi}f'(r)\right|_{r=r_{+}},
\ee
where $\kappa$ is the surface gravity. In fact, it is not necessary to substitute the expression $m(r)$ in Eq.(\ref{mr1}) into the above equation.
According to Eq.(\ref{fmr}), we know that $m(r_{+})=r_{+}/2$ at the event horizon. Moreover, using the field equation, Eq.(\ref{eom}), one can easily obtain
\be\label{tl}
T_{h}=\d{1}{4\pi r_{+}}-\d{r_{+}\mathcal{L}}{8\pi}.
\ee
We also use the new variables $x,Q$ to express the Hawking temperature, which is
\be
T_{h}=\frac{1024 Q^8 x_+^4-1024 Q^6 x_+^2+\pi ^4}{4096 \pi  q Q^9 x_+^5+4 \pi ^5 q Q x_+}
\ee
In Fig.\ref{fig2}, we show the temperatures of
 the regular black hole. It is shown that the black hole will never evaporate completely and radiate until the temperature becomes zero at some critical value of the event horizon. After that, the temperature is negative, which is physically unacceptable\footnote{Although there are some physically acceptable systems which have negative temperature, for black holes the same explanation does not work.}. The zero-temperature black hole is in fact the extremal one. Here one
 thing should be clarified. For the metric function depicted in Fig.\ref{fig1} the extremal case corresponds to fixed $x=x_e$ and $Q=Q_e$ because the metric is the function of two variables $x,~Q$ only. While the Hawking temperature depicted in Fig.\ref{fig2} is dependent on three variables $x_{+},~q,~Q$. Thus, choosing different $q,~Q$, the black hole always has the extremal case.

\begin{figure}[!htbp]
\includegraphics[width=7cm,keepaspectratio]{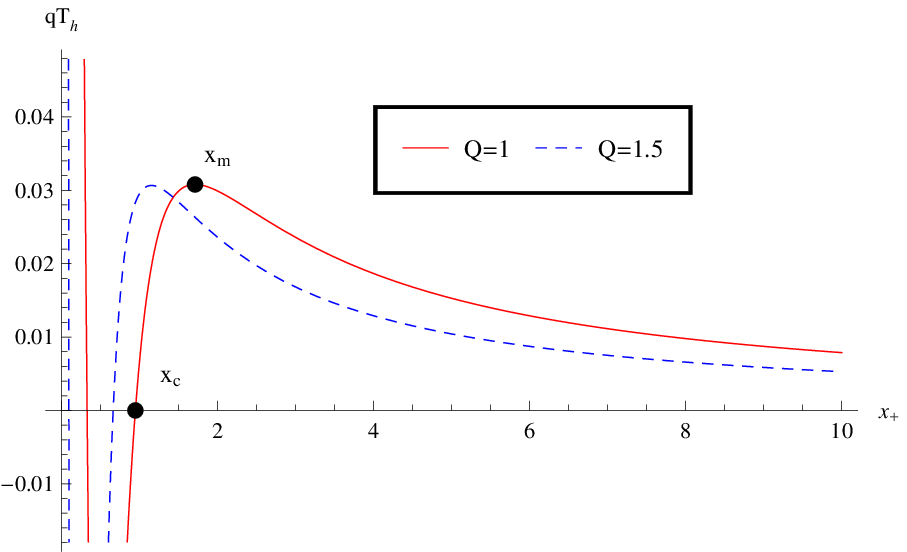}\hspace{0.5cm}
\includegraphics[width=7cm,keepaspectratio]{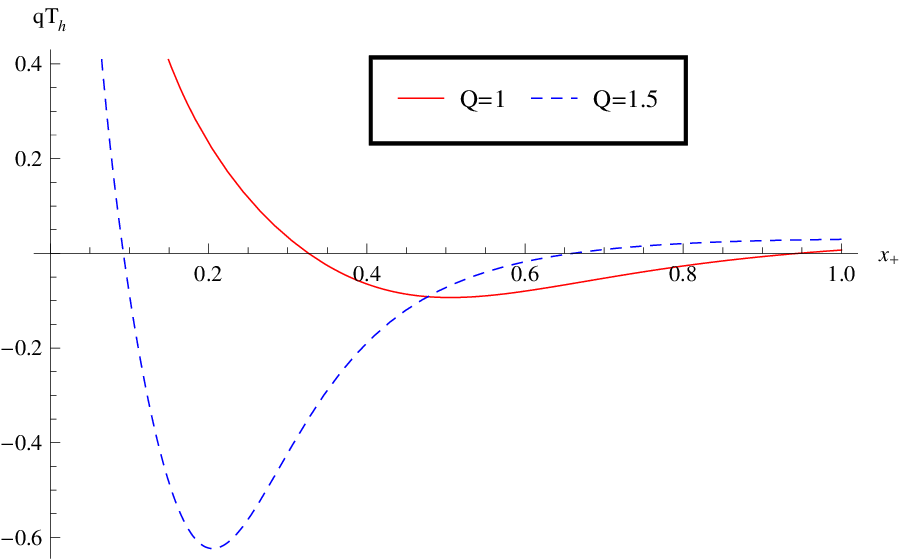}
\caption{$qT_h$ as functions  of  $x_{+}$ for $Q=1,~Q=1.5$. The subfigure on the RHS is a magnification of the subfigure on the LHS in the range $x\in[0,1]$.
 \label{fig2}
}
\end{figure}

Now we should discuss other thermodynamic quantities of the regular black hole, such as internal energy and entropy and heat capacity.
Entropy,  Komar energy, Smarr formula, phase transition, etc. of  regular black holes have been studied extensively
\cite{Cheng,Sharif,Myung1,Myung2,Myung3,Rabin:2008,Rabin:2009,Rabin:2011,Nicolini:2011,Smailagic,Dym3}. However, the definitions of the internal energy and the entropy of regular
black holes has some controversies\cite{CQG-ma}. Because our black hole solution is derived in general relativity, we insist that its entropy should satisfy the Bekenstein-Hawking area law, namely $S=A_{h}/4=\pi r_{+}^2$. Generally, for regular black holes, if one takes the black hole entropy as the area-law form, the black hole mass $M$ is no longer the internal energy of the system and thus the usual first law of black hole thermodynamics may violate\cite{CQG-ma}. However, black hole is a thermodynamic system, the first law of thermodynamics must be satisfied. Therefore, there must be some internal energy $U$ which fulfill the first law: $dU=TdS+\phi dq$, where $q$ is the magnetic charge and $\phi$ is the conjugate potential.
To understand the local stability of the regular black hole, we can calculate its heat capacity.
Although we do not know the detail form of the internal energy, we can still define the heat capacity at constant charge as
\be
C_q=\d{\p U}{\p T}|_{q}=T\d{\p S}{\p T}|_q=-\frac{2 \pi  q^2 Q^2 x_+^2 \left(1024 Q^8 x_+^4+\pi ^4\right) \left(1024 Q^8 x_+^4-1024 Q^6 x_+^2+\pi ^4\right)}{1048576 Q^{16} x_+^8-3145728 Q^{14} x_+^6+2048 \pi ^4 Q^8 x_+^4+1024 \pi ^4 Q^6 x_+^2+\pi ^8}.
\ee

\begin{figure}[!htbp]
\includegraphics[width=8cm,keepaspectratio]{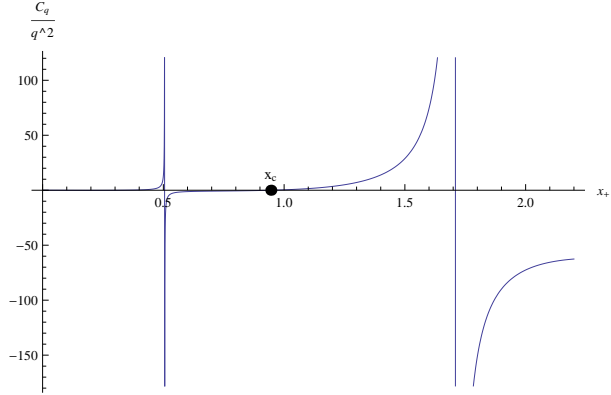}
\caption{$C_q/q^2$ as functions  of  $x_{+}$ for $Q=1$.
 \label{figcq}
}
\end{figure}

As is shown clearly in Fig.\ref{figcq} that $C_q$ will approach zero as the regular black
hole tends to the extremal one and diverge at the point where the
temperature of the black hole takes the maximum. According to Davies' viewpoint, this means that there will be
a phase transition at $x_+=x_{m}=1.71$ with fixed $Q=1$ for the black hole. The $C_q$ curve on the left of the point $x=x_c$, thus certainly the divergent point at smaller $x_{+}$, is meaningless because the temperature there is negative. The heat capacity is negative for larger black hole with $x_{+}>x_m$ and positive for smaller black hole with $x_c<x_{+}<x_m$. It means that the smaller black hole is thermodynamically stable and there may be a larger black hole/smaller black hole phase transition. At the critical point $x_m$, a second-order phase transition happens.

In addition, we can consider the thermodynamic quantities of the regular black hole from another perspective, namely the horizon thermodynamics proposed by Padmanabhan\cite{Pa2}. For a static, spherically symmetric spacetime the Einstein's
equations can be interpreted as the thermodynamic identity
\be\label{pv1}
dE=TdS-PdV,
\ee
where $T=T_h$, $S=A_{h}/4$, $E=\sqrt{A_{h}/16\pi}$,  $V=4\pi r_h^3/3$ , and for our action $P=-T_{1}^{~1}(r_h)/8\pi$.

One cannot define heat capacity  as $C=dE/dT$ directly as is done in\cite{Dymnikova}. Firstly, due to the presence of matter fields, $P\neq 0$, thus the identity Eq.(\ref{pv1}) cannot return to the usual first law of black hole thermodynamics $dM=TdS$. Secondly, even in vacuum case, for many regular black holes $dM\neq TdS$ with the area-law entropy\cite{CQG-ma}. Thirdly, in the framework of horizon thermodynamics one cannot define the heat capacity at constant volume $C_V$ because that means constant $r_{+}$. Therefore, we introduce the enthalpy $H=E+PV$ of the system to define heat capacity at constant pressure as
\be
C_P=\left.\d{\p H}{\p T}\right|_{P}=\d{\left.\d{\p H}{\p r_{+}}\right|_{P}}{\left.\d{\p T}{\p r_{+}}\right|_{P}}.
\ee
According to Eq.(\ref{tl}), one can easily obtain
\be\label{CP}
C_{P}=\frac{2 \pi  r_+^2 \left(8 \pi  P r_+^2+1\right)}{8 \pi  P r_+^2-1}.
\ee
Obviously, this result do not rely on the concrete black hole solution. It only depends on the gravitational theory under consideration. For general relativity, the heat capacity at constant pressure is given by Eq.(\ref{CP}), while for other theories of gravity it has different forms\cite{pvthermo}.
For our NED model,
\be
P=-\d{\mathcal{L}}{16\pi}=\frac{-128 M^4 q^2}{1024 \pi  M^4 r_{+}^4+\pi ^5 q^8}\leq 0,
\ee

\begin{figure}[!htbp]
\includegraphics[width=7cm,keepaspectratio]{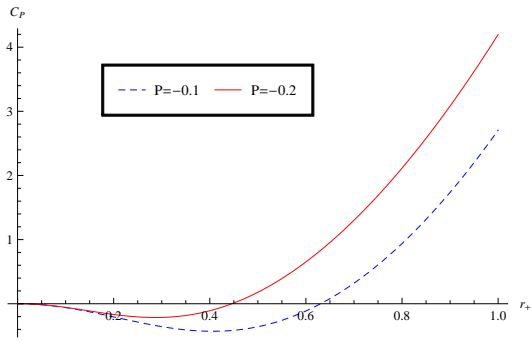}\hspace{0.5cm}
\caption{The heat capacity at constant pressure $C_{P}$ as function  of  $r_{+}$ for different pressure.\label{CP}
}
\end{figure}
Obviously the behaviors of $C_P$ and $C_q$ are very different. This is because the two heat capacities are derived in two different frameworks of black hole thermodynamics. $C_P$ does not diverge. However, there is a point after that $C_P$ is positive which means that the black holes in this case is local stable.
As we have noted above, this conclusion applies to all kinds of black holes derived in general relativity, not only our regular black hole.

\par

\section{Concluding remarks}

In this paper, we considered a model of nonlinear electrodynamics. We expect that the model is physically allowable because of three reasons. Firstly, the NED model returns back to the Maxwell electrodynamics in the weal field limit. Secondly, the NED model can produce the phenomenon of vacuum birefringence which is a kind of nonlinear effect originated from the nonlinearity of the NED. Thirdly, the electromagnetic field in the NED model satisfy the weak energy condition which means the energy density of the matter fields measured by a local observer is nonnegative. By coupling the NED to general relativity, we obtain a magnetically charged regular black hole only when the parameter $\beta$ in the NED model satisfies the condition $\beta=\left(\frac{\pi  q^{3/2}}{8 M}\right)^4$. We also shown that only when $M\geq0.9893q$ the regular black hole exists with one or two horizons. When $M<0.9893q$ we only have a particle-like solution.

We also discuss the thermodynamic quantities and thermodynamic properties of the regular black hole. From Fig.\ref{fig2} one can see that the Hawking temperature becomes zero as the black hole tends to the extremal one. After that, the temperature is negative, which we think it is meaningless for black hole system.  The regular black hole may have the similar problems to those we have studied in \cite{CQG-ma}. To discuss the local thermodynamic stability of the regular black hole, we calculate the heat capacity at constant magnetic charge $C_q$ according to the corrected first law of black hole thermodynamics. It is shown that $C_q$ is negative for larger black hole and positive for smaller one, which means that the smaller black hole is more stable. At the point where the temperature takes maximum, $C_q$ will diverge. According to Davies's viewpoint, a second-order phase transition happens there.

In addition, we investigate the thermodynamics of the regular black hole according to Padmanabhan's horizon thermodynamics. Dymnikova has ever employed this method to discuss thermodynamics of regular black holes\cite{Dymnikova}. However, the heat capacity she defined is problematic. In fact, in the framework of horizon thermodynamics one can only define heat capacity at constant pressure $C_P$. Furthermore, the $C_P$ is only dependent on the gravitational theories,  not the concrete black hole solutions derived in the gravitational theories. In our model, the pressure is always negative so long as the mass or magnetic charge are nonzero. As is shown in Fig.\ref{CP}, the behavior of $C_P$ is very different from that of $C_q$ as one expected.

\section*{Acknowledgements}
The author would like to thank Professor Ren Zhao for fruitful discussions and Professor S. Kruglov for useful communication.
This work is supported in part by the National Natural Science Foundation of China (NSFC) under Grants
Nos.11475108,11175109 and by the Doctoral Sustentation
Fund of Shanxi Datong University (2011-B-03).

\end{document}